\documentclass[a4paper,11pt]{article}
\usepackage{fancyheadings}
\usepackage[dvips]{graphicx}
\usepackage{ifthen}
\usepackage{amssymb}
\usepackage{epsfig}
\usepackage[errorshow]{tracefnt}
\usepackage{hyperref}
\newcommand\ffam{\sffamily}
\newcommand\fser{\bfseries}
\newcommand\fsh{\upshape}

\newcommand\scs{\scriptstyle}

\newcommand\npb{\nopagebreak}

\newcommand\benu{\begin{enumerate}}
\newcommand\eenu{\end{enumerate}}
\newcommand\bit{\begin{itemize}}
\newcommand\eit{\end{itemize}}

\newcommand{\be}{\begin{eqnarray}}
\newcommand{\ee}{\end{eqnarray}}
\newcommand{\bd}{\begin{displaymath}}
\newcommand{\ed}{\end{displaymath}}
\newcommand{\bq}{\begin{equation}}
\newcommand{\eq}{\end{equation}}

\newcommand\ta{{\tt a}}
\newcommand\tb{{\tt b}}
\newcommand\tc{{\tt c}}
\newcommand\td{{\tt d}}

\newcommand\ie{{\it i.e.,}}
\newcommand\eg{{\it e.g.,} }

\newcommand\srule{\noindent\npb \vspace{12pt}\rule[-4pt]{.4pt}{4pt}\hrulefill \rule[-4pt]{.4pt}{4pt}\\}
\newcommand\erule{\vspace{4pt}\npb\\ \rule{.4pt}{4pt}\hrulefill \rule{.4pt}{4pt}\vspace{5pt}\\}
\newcommand\ind{\hspace*{1cm}}
\newcommand\lskip{\vspace{16pt}}
\newcommand\sk{\vspace{9pt}}

\newcommand\nc{$\raisebox{2pt}{$\scs \star\star$}$}

\begin{document}

%
%

\thispagestyle{empty}

\null\vskip-0.5cm

\begin{tabular}{l}
\epsfig{file=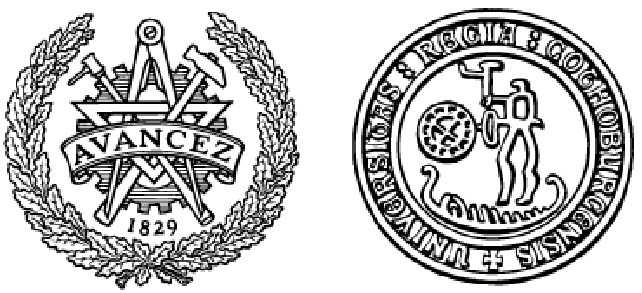,height=1.7cm}
\end{tabular}
\hfill
\begin{tabular}{r}
G\"oteborg ITP preprint\\
{\tt hep-th/0105086}\\
May 10, 2001
\end{tabular}

\hrulefill
\vskip2cm

\begin{center}
{\upshape\sffamily\Large GAMMA: A Mathematica package for performing
$\Gamma$-matrix algebra and Fierz transformations in arbitrary dimensions} \\[4mm]
\end{center}

\vspace*{4mm}
\begin{center}
    {\ffam\fsh\large Ulf Gran\footnote{E-mail: gran@fy.chalmers.se}}\\
\end{center}

\vspace{1cm}
\begin{center}
    {\ffam\fsh Department of Theoretical Physics\\*[1mm]
    Chalmers University of Technology and G\"oteborg University\\*[1mm]SE-412 96 G\"oteborg, Sweden}
\end{center}
\normalsize

%
%

\vspace{1cm}
\centerline{\ffam\fser Abstract}
\vspace{2mm}

\noindent We have developed a Mathematica package capable of performing $\Gamma$-matrix
algebra in arbitrary (integer) dimensions.
As an application we can compute Fierz transformations.  

%
%
\newpage

\begin{center}
\Large COPYRIGHT AND DISCLAIMER OF WARRANTY\vspace{1cm}
\end{center}

\noindent Copyright \copyright\ 2001 Ulf Gran.\lskip

\noindent This documentation is part of the GAMMA package, available for
download at {\tt http://fy.chalmers.se/\~{}gran/}, developed for Mathematica
version 4.0.\lskip

\noindent The GAMMA package is distributed in the hope that it will be useful but WITHOUT
ANY WARRANTY. Neither the author nor any distributor accepts responsibility to anyone for the
consequences of using it or for whether it serves any particular purpose or
works at all, unless he says so in writing.\lskip

\noindent Permission is granted for use and non-profit distribution of this
package, in its original form,
providing that this notice be clearly maintained, but the right to distribute
this package for profit or as part of any commercial product is specifically
reserved for the author. The same permission and reservation applies to the
formatted document distributed with this package.\lskip

\noindent If you find the GAMMA package useful for doing research, acknowledging the author's work by proper
reference to the package would be appreciated.

%
%

\newpage
\section{Introduction}

The motivation to develop this package arose from calculations in
eleven-dimensional supergravity \cite{Cederwall:2000,Cederwall:2000:2}, which
required an immense amount of both pure $\Gamma$-matrix algebra and Fierz
transformations to be performed. 
The rule-based programming of Mathematica \cite{Wolfram:1988} is ideally suited for handling
algebraic computations, compared to ordinary procedural programming. For a very
nice comparison between rule-based and procedural programming, as well as a toy
implementation of $\Gamma$-matrix algebra in four dimensions, see
\cite{Jamin:1993}, which describes a Mathematica package performing $\Gamma$-matrix
algebra in $(4-\epsilon)$ dimensions.
This was the first Mathematica
implementation of $\Gamma$-matrix algebra according to the 't Hooft-Veltman scheme. There are quite a few programs
handling $\Gamma$-matrices in four-dimensions
\cite{Strubbe:1974,Hearn:1987,Krawczyk:1987,Vermaseren:1989,Hsieh,Kublbeck:1990,
Mertig:1991}, intended to be used when
computing Feynman diagrams.
There is also a Mathematica package handling computations in $D=10$, $N=1$
supergravity, including $\Gamma$-matrix algebra, \cite{Saulina:1995}.

In section \ref{algebra}, we will introduce and explain the functions in the
package concerning
the algebraic manipulation of $\Gamma$-matrices. This will be done through
simple examples. In section
\ref{fierz} we will turn to the functions concerning fierzing. 
A complete list of the functions in the package, including brief descriptions,
can be found in appendix \ref{list}. In appendix \ref{redef} we list the
built-in Mathematica functions that are redefined by the package.
GAMMA is available for download at {\tt http://fy.chalmers.se/\~{}gran/}.

\section{$\Gamma$-matrix algebra}\label{algebra}

To load the package type

\srule
{\tt \ind In[1]:=<<GAMMA.m}
\erule
and remember to include the path if the package is put in a directory where Mathematica
does not automatically search.
By default, the space-time dimension is set to 11 and the spinor dimension is
set to 32.  
If you want to use any other values they must be specified before any
calculations can be done. This is done through, \eg\

\srule
{\tt \ind In[1]:=SetDim[10]; SetSpinorDim[16];}
\erule
The spinor dimension is used only when calculating the trace of $\Gamma$-matrices.
We now introduce the various objects that can be handled. 
Kronecker deltas, $\Gamma$-matrices, tensors and tensor-spinors are entered like 

\srule
{\tt \ind In[1]:=Delta[\{a,b\},\{c,d\}]\\
\ind Out[1]:=$\delta^{\tt a b}_{\tt c d}$\sk\\
\ind In[2]:=GammaProd[\{a,b\},\{c\}]\\
\ind Out[2]=$\Gamma^{\tt ab}\nc\Gamma^{\tt c}$\sk\\
\ind In[3]:=Tensor[X,\{a,b,c\}]\\
\ind Out[3]=$\tt X_{\tt a b c}$\sk\\
\ind In[4]:=TensorSpinor[Y,\{a,b,c\}]\\
\ind Out[4]=$\tilde{\tt Y}_{\tt a b c}$
}\erule
The indices in each list in these expressions are assumed to be antisymmetric
and will be put in a canonical order (defined by Mathematica's sort algorithm,
which therefore means alphabetical order), sometimes resulting in a sign in front of
the canonically ordered expression.
In order to discern tensors from tensor-spinors (since we do not use explicit
spinor indices) we have added a tilde over the
symbol designating the tensor-spinor. This indicator can be
changed by, \eg\

\srule
{\tt \ind In[1]:=SetTensorSpinorInd[OverHat]}
\erule
in which case the Mathematica function {\tt OverHat} is applied to the symbol
designating the tensor-spinor. See the Mathematica
documentation for a complete list of possible choices. It is also possible to
change the formatting of the symbols designating a tensor, where the
default function is

\srule
{\tt \ind In[1]:=SetTensorInd[Identity]}
\erule
Since all calculations are assumed to be performed in a Lorentz frame, we do not
have to care about whether an index is upstairs or downstairs and
contracted indices may both be upstairs or downstairs.
Enforcing explicit antisymmetry in some indices in an expression like
$\Gamma_{\tt [a} {\tt X}_{\tt b]}$ is done by

\srule
{\tt \ind In[1]:=ASym[GammaProd[\{a\}] Tensor[X,\{b\}],\{a,b\}]\\
\ind Out[1]=$\tt\frac{1}{2}(X_b\Gamma_a-X_a\Gamma_b)$}
\erule
Using the function {\tt Sym[]} instead enforces symmetry in the specified indices.

We now turn to the functions concerning the manipulation of
$\Gamma$-matrices. We have already seen how to write a product of
$\Gamma$-matrices, if we want to expand a general product we write

\enlargethispage*{10pt}
\srule
{\tt \ind In[1]:=GammaExpand[GammaProd[\{a,b\},\{c\},\{d\}]]\\
\ind
Out[1]=$\delta^\tb_\tc\Gamma_\ta\nc\Gamma_\td-\delta^\ta_\tc\Gamma_\tb\nc\Gamma_\td+\Gamma_{\tt
abc}\nc\Gamma_\td$}\
\erule
where the product of the first two $\Gamma$-matrices have been expanded. Since
only the product of two $\Gamma$-matrices are expanded each time {\tt GammaExpand[]}
is applied, repeated application is sometimes needed.
In the expansion we only use the algebra $\{\Gamma^a,\Gamma_b\}=2\delta^a_b$
and therefore, as long as the Kronecker deltas have an equal number of upstairs
and downstairs indices, there will be no difference between various
signatures. In a specific expression, if we want a Kronecker delta to have, \eg
only upstairs (or downstairs) indices we get a trivial difference since using
Minkowski signature the Kronecker delta becomes an $\eta$ while it stays a delta in the euclidean
case. Therefore, with the understanding that a {\tt DeltaProd[]} has an equal
number of upstairs and downstairs indices, the expressions will look the same
for all signatures.  

Since we know that the
expression we started with in this example above is antisymmetric in \ta\ and \tb\ we
can simplify the result by putting these indices in canonical order by

\srule
{\tt \ind In[2]:=ACanonicalOrder[\%,\{a,b\}]\\
\ind Out[2]=$-2\delta^\ta_\tc \Gamma_\tb\nc\Gamma_\td+\Gamma_{\tt abc}\nc\Gamma_\td$}
\erule
remembering that that the expression is antisymmetric in \ta\ and \tb. 
Even though the simplification in this simple case is not that significant the
importance to use this function, when applicable, in more complicated cases can not be underestimated. 
The function {\tt SCanonicalOrder[]} puts indices in canonical order assuming
symmetry instead. When indices are contracted between adjacent $\Gamma$-matrices
it is more efficient to use

\srule
{\tt \ind In[1]:=GammaContract[GammaProd[\{a,b,c\},\{b,c\}]]\\
\ind Out[1]=-90$\Gamma_\ta$}
\erule
which is specialized for this situation, than to use {\tt GammaExpand[]}. To
compute traces of expressions involving $\Gamma$-matrices we write\footnote{When
tracing a $\Gamma$-matrix with the same number of indices as the spacetime
dimension, when this number is odd, {\tt GammaProd[]} is kept and the user can himself
replace it with a Levi-Civita tensor density according to his conventions (note
that {\tt GammaProd[]} is
{\em not} multiplied by the spinor dimension).}

\srule
{\tt \ind In[1]:=GammaTrace[GammaProd[\{a,c\},\{b,c\}]]\\
\ind Out[1]=-320$\delta^\ta_\tb$}
\erule  
When using tensor-spinors, which are assumed to be irreducible and therefore
$\Gamma$-traceless, it is sometimes needed to rearrange the $\Gamma$-matrices in
order to be able to use the $\Gamma$-tracelessness to simplify an
expression. We see in this example

\srule
{\tt \ind In[1]:=GammaProd[\{a,b,c\}]\nc TensorSpinor[Y,\{c\}]\\
\ind Out[1]=$\Gamma_{\tt abc}\nc\tilde{\tt Y}_\tc$\sk\\
\ind In[2]:=GammaExtract[\%,\{a,b\}]\\
\ind Out[2]=-2$\Gamma_\ta\nc\tilde{\tt Y}_\tb$}
\erule
that we have to use {\tt GammaExtract[]} to be able to simplify the
expression. What {\tt GammaExtract[]} does is to extract one $\Gamma$-matrix at
a time, having an index contracted to the tensor-spinor, from, in this case, the
$\Gamma_{\tt abc}$ term, in this way enabling the use of $\Gamma$-tracelessness, \eg\ $\Gamma_\tc
\tilde{\tt Y}_\tc=0$. 
Note that it is very important to use \nc, \ie\ {\tt NonCommutativeMultiply[]},
in this example since there are more than one object carrying spinor indices.
After some manipulations, one often ends up with expressions where contracted,
\ie\ dummy,
indices are denoted by different symbols in otherwize identical expressions like in

\srule
{\tt \ind Out[1]=${\tt X}_\ta \Gamma_\ta+{\tt X}_\tb \Gamma_\tb$}
\erule
In this case Mathematica can not tell that these terms are identical and we have to
rename the dummy indices using some canonical set of indices. The is done by

\srule
{\tt \ind In[2]:={\tt RenameDummy}[\%]\\
\ind Out[2]=2${\tt X}_{\tt d1}\Gamma_{\tt d1}$}
\erule 
The set of indices ${\tt \{d1,d2,\ldots\}}$ are used to denote dummy indices and
are therefore preferably avoided. It is possible to cut, paste and copy
formatted expressions containing $\Gamma$-matrices and also to enter expressions
as they look in the output form. Converting expressions to \TeX\ format is done by {\tt TeXForm[expr]}.

\section{Fierz transformations}\label{fierz}

We can also use GAMMA to perform Fierz transformations. This will be illustrated
using a simple example in which the free vector indices of the
$\Gamma$-matrices are contracted to a tensor. Start by making the definitions

\srule
{\tt \ind structures=\{x1,x2,y1,y2,y3,z1,z2,z3\};\sk\\
\ind ind=HoldForm[$\tt\frac{1}{3}tr[P\nc\tt X]\nc\tt Q+\tt\frac{2}{3}P\nc\tt
X\nc \tt Q$];\sk\\
\ind x1=\{\{a\},\{b\},ind,\{\{\},\{\}\}\};\\
\ind x2=\{\{c1\},\{c1,a,b\},ind,\{\{\},\{\}\}\};\sk\\ 
\ind y1=\{\{a,b\},\{\},ind,\{\{\},\{\}\}\};\\
\ind y2=\{\{a,c1\},\{b,c1\},ind,\{\{\},\{\}\}\};\\
\ind y3=\{\{c1,c2\},\{a,b,c1,c2\},ind,\{\{\},\{\}\}\};\sk\\
\ind z1=\{\{a,b,c1,c2,c3\},\{c1,c2,c3\},ind,\{\{\},\{\}\}\};\\
\ind z2=\{\{a,c1,c2,c3,c4\},\{b,c1,c2,c3,c4\},ind,\{\{\},\{\}\}\};\\
\ind z3=\{\{c1,c2,c3,c4,c5\},\{a,b,c1,c2,c3,c4,c5\},ind,\{\{\},\{\}\}\};
}
\erule
The list {\tt structures} contains all the different terms we want to
relate by fierzing. A term {\tt w1=\{\{a\},\{b\},ind,\{\{c\},\{d\}\}\}}
corresponds to
\begin{equation}
(\Gamma^a)_{\alpha\beta}(\Gamma^b)_{\gamma\delta}\delta^c_d\,. 
\end{equation}
Note that the terms in this example have two free vector indices, \ta\ and \tb.
To perform the Fierz transformation we contract two spinor indices of each term
with a basis of $\Gamma$-matrices, $\Gamma^{(1)}$, $\Gamma^{(2)}$ and
$\Gamma^{(5)}$ for symmetric spinor indices and $C$,
$\Gamma^{(3)}$ and $\Gamma^{(4)}$ for antisymmetric spinor indices.
In our case we are interested in three symmetric indices, $(\alpha\beta\gamma)\delta$.
Since we do use explicit spinor indices on the $\Gamma$-matrices we use the
expression {\tt ind} to specify the symmetry of the four spinor 
indices. In {\tt ind}, {\tt P} will be replaced by the first $\Gamma$-matrix in a
term and {\tt Q} by the second. {\tt X} will be replaced by the basis matrices
and {\tt tr} by the function {\tt GammaTrace}. It is then easy to see that {\tt
ind} corresponds to the symmetry specified above\footnote{Contract a basis
$\Gamma$-matrix, $\Gamma^{ X}$, with two of the symmetrized indices,
$(\Gamma^{ X})^{\alpha\beta}\Gamma^{ P}_{(\alpha\beta}\Gamma^{
Q}_{\gamma)\delta}=\frac{1}{3}{\mathrm tr}(\Gamma^{X}\Gamma^{P})\Gamma^{
Q}_{\gamma\delta}+\frac{2}{3}(\Gamma^{ P}\Gamma^{X}\Gamma^{Q})_{\gamma\delta}$,
where the RHS is the expression to be used as {\tt ind}.}. We are now ready to perform
the Fierz transformation,

\renewcommand{\arraystretch}{1.5}
\srule
{\tt \ind
In[1]:=FierzSolve[Fierz[\{y1,y2,y3\},structures,\{a,b\},S,\\\ind\ind \ind MakeTensor]]\\
\ind Out[1]=\begin{eqnarray*}\left(\begin{array}{rrr}
-\frac{4}{5}&\frac{3}{5}&-\frac{1}{5}\\
\frac{9}{5}&-\frac{8}{5}&\frac{1}{5}\\
\frac{24}{5}&-\frac{48}{5}&\frac{11}{5}\\
-\frac{36}{5}&\frac{2}{5}&\frac{1}{5}\\
18&-8&3
\end{array}\right)\end{eqnarray*}}
\erule
where the first argument to {\tt Fierz} is the terms that are chosen as the
basis in which the other terms will be expressed. It is important to know
exactly how many linearly independent terms there are and to get this
information we recommend the lie algebra program Lie \cite{Cohen:1998}.
The second argument is the list of terms, the third is a list of the free vector
indices, the fourth is S since we are fierzing symmetric indices in this example
(use A for antisymmetric indices). The last argument {\tt MakeTensor} is used
to indicate that the symmetry of the free indices is that of a tensor. 
In the output, each row contains the coefficients for a term in the chosen
basis. In this case, \eg\ ${\tt z2=-\frac{36}{5}y1+\frac{2}{5}y2+\frac{1}{5}y3}$.

When we Fierz tensor-spinors, care has to be taken in order to be able to use the
$\Gamma$-tracelessness. If, \eg\ the index {\tt b} in the previous example is
contracted to a vector-spinor having the spinor index $\gamma$ we would have to
use

\srule
{\tt \ind ind=HoldForm[$\tt\frac{1}{3}tr[P\nc\tt X]\nc\tt Q\nc \tt
TensorSpinor[\tt Y,\{\tb\}]\\\ind\ind+\tt\frac{2}{3}P\nc\tt
X\nc \tt Q\nc \tt TensorSpinor[\tt Y,\{\tb\}]$]}
\erule
The tensor-spinor is automatically removed after the simplification is complete when fierzing.

Fierz transformations like the one above where the free vector indices are
contracted to an antisymmetric tensor are relatively simple to perform, it is harder to Fierz
hook-tensors\footnote{A hook-tensor is denoted $Z_{c_1\ldots c_p,a}$ satisfying
$Z_{[c_1\ldots c_n,a]}=0$ and $Z_{c_1\ldots c_{n-1} c}{}^c=0$.}.  
Even though GAMMA may not be able to directly calculate the Fierz transformation as
shown above, it can still be of great use. 
We can try to use {\tt Fierz[]} but if that fails we can use {\tt
FierzTerm[]} instead, which contracts the basis of $\Gamma$-matrices to a specific term, or list of terms, simplifies and
then returns the result. The problem with hook-tensors is that a lot of
Kronecker delta functions are generated and it is not easy to rewrite them on a
canonical form allowing Mathematica to determine which terms are linearly
independent.

The way to specify a hook-tensor term is a bit different from the tensor
case. Let us look at a term arising when we want to Fierz a (11000)
representation,
\begin{equation}
(\Gamma^{c_1 c_2})_{\alpha\beta}(\Gamma^{f a})_{\gamma\delta}Z_{c_1 c_2,a}\,,\label{hf}
\end{equation}
where we have one free vector index. This term is represented by 

\srule
{\tt \ind f1=\{\{\},\{f\},\{c1,c2\},\{a\},2,ind,\{\{\},\{\}\}\};}
\erule
Here, the first entry corresponds to the vector indices on the first $\Gamma$-matrix, not
including indices contracted to the hook-tensor, and the second entry is defined
in the same way
but for the second $\Gamma$-matrix. The third entry is the antisymmetric
hook-tensor indices, except those that sit on a Kronecker delta
function (after
removing the hook-tensor\footnote{For a term as in (\ref{hf}) we have to introduce a Kronecker delta function
when removing the hook-tensor if the free index {\tt f} would sit on the
hook-tensor instead of on a $\Gamma$-matrix.}), and the fourth entry is
the symmetric index on the hook-tensor. The fifth entry is the number of vector
indices on the first $\Gamma$-matrix, {\tt ind} again specifies the structure of
the spinor indices and the last entry is the potential Kronecker delta function indices.
When we remove the hook-tensor in (\ref{hf}) we must enforce the hook-tensor
symmetries on the $\Gamma$-matrix terms, this is done by the function {\tt
MakeHook[term,gmatrix]}, which at the same time contracts two spinor indices to
the spinor indices on the $\Gamma$-matrix {\tt gmatrix} as specified by {\tt ind}.
By writing 

\srule
{\tt \ind In[1]:=FierzTerm[\{f1\},\{f\},S,MakeHook]}
\erule
we generate the expressions obtained by contracting the symmetric $\Gamma$-matrix
basis, in this case, to the hook structure specified by {\tt f1}, treating the
indices contained in the list of the second argument as antisymmetric. These
expressions can then be studied ``by hand'' and in this way we can also simplify 
hook fierzes.

\section{Summary}\label{summary}
We have presented a Mathematica package capable of performing $\Gamma$-matrix
algebra in arbitrary (integer) dimensions. The various functions have been
introduced through examples and a complete list of all functions in the package,
including brief descriptions, can be found in appendix \ref{list}. As a specific
application, we have considered Fierz transformations, which in general require
a lot of $\Gamma$-matrix calculations.
Note that some of the built-in functions in Mathematica have been altered, see
appendix \ref{redef} for a complete list.

Comments, suggestions for improvements and reports regarding any discovered bugs
are appreciated. When reporting bugs, please include a Mathematica notebook
illustrating the bug in question. Reporting bugs in this way will result in the
quickest possible bug-fixes. 
Please note that only the most basic error handling is implemented. Improved
error handling invariably results in a slower program and in the choice between
error handling or speed, we chose the latter. This unfortunately means that what
might at first sight look like a bug quite often can be attributed to some error
in the input. 

\vspace{5mm}
\noindent{\Large {\bf Acknowledgments}}
\vspace{5mm}\\
We would like to thank Martin Cederwall, Mikkel Nielsen and Bengt E.W. Nilsson
for discussions and collaborations leading to the construction of GAMMA.  
We would also like to thank M\'{a}ximo Ba\~{n}ados, Mikkel Nielsen, Dimitrios Tsimpis and
Anders Westerberg for valuable comments regarding the manuscript and the program. 

\appendix
\section{List of commands}\label{list}
Here is a list of all the user-available functions in the GAMMA package together
with a briefs descriptions.
\begin{description}
\item[{\tt ACanonicalOrder[expr,list]}] assumes that {\tt expr} is antisymmetric
in the indices given by {\tt list} and puts them in canonical order, thus
enabling additional simplification.
\item[{\tt ASym[expr,list]}] explicitly antisymmetrizes {\tt expr} in the
indices contained in {\tt list}.
\item[{\tt Delta[list1,list2]}] represents a Kronecker delta with the two groups
of indices given by {\tt list1} and {\tt list2}.
\item[{\tt Fierz[basislist,structures,asymlist,sym,function]}] $\!\!$performs the
Fierz transformation of the terms specified in {\tt structures} using the terms
in the list {\tt basislist} as basis elements. Antisymmetry is assumed in the
indices contained in the list {\tt asymlist} and {\tt sym} ({\tt A} or
{\tt S}) gives the symmetry of the spinor indices to be fierzed. Finally, {\tt
function} is applied to generate explicit $\Gamma$-matrix expressions given a
term in {\tt structures}.
\item[{\tt FierzSolve[eqlist]}] solves the list of equations generated by {\tt Fierz[]}.
\item[{\tt FierzTerm[termlist,asymlist,sym,function]}] as {\tt Fierz[]} but
does not perform the complete Fierz transformation. Instead it produces a list
of expressions obtained by contracting two of the spinor indices, as specified
by {\tt ind}, with the basis of $\Gamma$-matrices specified by
{\tt sym}. 
\item[{\tt GammaContract[expr]}] simplifies the expression {\tt expr} containing
contracted indices on adjacent $\Gamma$-matrices.
\item[{\tt GammaExpand[expr]}] expands the product of the first two $\Gamma$-matrices in
{\tt expr}.
\item[{\tt GammaExtract[expr,list]}] extracts $\Gamma$-matrices that contract
indices on a tensor-spinor, assuming antisymmetry in the indices contained in {\tt
list}, in order to be able to invoke the $\Gamma$-tracelessness of the tensor-spinor.
\item[{\tt GammaProd[list1,...,listn]}] represents the product of {\tt n}
$\Gamma$-matrices with indices given by {\tt list1} to {\tt listn}.
\item[{\tt GammaTrace[expr,list]}] computes the $\Gamma$-trace of an expression
{\tt expr}
containing $\Gamma$-matrices, enforcing antisymmetry in the indices contained in
{\tt list}.
\item[{\tt MakeHook[term,basiselement]}] makes an explicit $\Gamma$-matrix
expression from {\tt term} contracting the $\Gamma$-matrix {\tt basiselement} with
two of the spinor indices, as specified by {\tt ind}.
\item[{\tt MakeTensor[term,basiselement]}] makes an explicit $\Gamma$-matrix
expression from {\tt term} contracting the $\Gamma$-matrix {\tt basiselement} with
two of the spinor indices, as specified by {\tt ind}. 
\item[{\tt RenameDummy[expr]}] renames dummy (contracted) indices in {\tt expr}, thus
enabling additional simplification.
\item[{\tt SCanonicalOrder[expr,list]}] assumes that {\tt expr} is symmetric
in the indices given by {\tt list} and puts them in canonical order, thus
enabling additional simplification.
\item[{\tt SetDim[dim]}] sets the space-time dimension to {\tt dim}.
\item[{\tt SetTensorInd[function]}] sets the Mathematica function to be
applied to the symbol designating a tensor to {\tt function}.
\item[{\tt SetTensorSpinorInd[function]}] sets the Mathematica function to be
applied to the symbol designating a tensor-spinor, in order to discern it from a tensor, to {\tt function}.
\item[{\tt SetSpinorDim[dim]}] sets the spinor dimension to {\tt dim}.
\item[{\tt Sym[expr,list]}] explicitly symmetrizes {\tt expr} in the
indices contained in {\tt list}.
\item[{\tt Tensor[sym,list]}] represents a tensor, denoted by the symbol {\tt sym},
with indices given by {\tt list}.
\item[{\tt TensorSpinor[sym,list]}] represents a tensor-spinor, denoted by the
symbol {\tt sym}, with (vector) indices given by {\tt list}.
\end{description}

\section{Redefined Mathematica built-in functions}\label{redef}
\begin{description}
\item[{\tt NonCommutativeMultiply[]:}] Redefining {\tt NonCommutativeMultiply[]}
is what enables us to use new, noncommutative, objects like $\Gamma$-matrixes and spinors
in Mathematica. {\tt
NonCommutativeMultiply[]} can also be entered as {\tt **}.
\item[{\tt Power[]:}] In order to be able to interpret expressions entered in the
output format, and also to be able to cut, copy and paste expressions, we have
to modify {\tt Power[]} so that \eg\ $\Gamma^\mu$ is replaced by {\tt GammaProd[{$\mu$}]}. 
\item[{\tt Subscript[]:}] The same reason as for {\tt Power[]} except that {\tt
Subscript[]} takes care of downstairs indices instead of upstair ones.
\end{description}

%
%
\cleardoublepage
\pagestyle{plain}
\def\href#1#2{#2}
\bibliographystyle{utphysm}
\bibliography{ugref}

\end{document}